\documentclass[number,citesort,dvips]{arxstspdf}
\usepackage{graphics,flushend}

\volume{24}
\issue{4}
\pubyear{2009}
\firstpage{561}
\lastpage{573}
\doi{10.1214/09-STS290}

\begin{document}
\begin{frontmatter}

\title{Replication in Genome-Wide Association Studies}
\runtitle{Replication in GWAS}

\begin{aug}
\author[A]{\fnms{Peter} \snm{Kraft}},
\author[B]{\fnms{Eleftheria} \snm{Zeggini}}
and
\author[C]{\fnms{John P. A.} \snm{Ioannidis}\ead[label=e3]{jioannid@cc.uoi.gr}\corref{}}
\runauthor{P. Kraft, E. Zeggini and J. P. A. Ioannidis}

\affiliation{Harvard School of Public Health, Wellcome Trust Sanger Institute and
Wellcome Trust Centre for Human Genetics, and University of Ioannina,
Biomedical Research Institute and Tufts University}

\address[A]{Peter Kraft is Associate Professor, Departments of Epidemiology and Biostatistics,
Harvard School of Public Health, Boston, Massachusetts, USA.}
\address[B]{Eleftheria Zeggini is Group Leader, Applied Statistical Genetics,
Wellcome Trust Sanger Institute, Hinxton, UK and
Wellcome Trust Centre for Human Genetics, University of Oxford, Oxford, UK.}
\address[C]{John P. A. Ioannidis is Professor and Chairman, Clinical and Molecular Epidemiology Unit,
Department of Hygiene and
Epidemiology, University of Ioannina School of Medicine, Ioannina, Greece,
Biomedical Research Institute, Foundation for Research and
Technology-Hellas, Ioannina, Greece and Director,
Tufts Clinical and Translational Science Institute and Center for
Genetic Epidemiology and Modeling, Tufts Medical Center and Tufts
University School of Medicine, Boston, Massachusetts, USA \printead{e3}.}
\end{aug}

\begin{abstract}
Replication helps ensure that a genotype-phenotype association observed
in a genome-wide association (GWA) study represents a credible
association and is not a chance finding or an artifact due to
uncontrolled biases. We discuss prerequisites for exact
replication, issues of heterogeneity, advantages and disadvantages of
different methods of data synthesis across multiple studies, frequentist
vs. Bayesian inferences for replication, and challenges that arise from
multi-team collaborations. While consistent replication can greatly
improve the credibility of a genotype-phenotype association, it may not
eliminate spurious associations due to biases shared by many studies.
Conversely, lack of replication in well-powered follow-up studies
usually invalidates the initially proposed association, although
occasionally it may point to differences in linkage disequilibrium or
effect modifiers across studies.
\end{abstract}


\end{frontmatter}

\section{Introduction} \label{sec:1}

Reproducibility has long been considered a key part of the scientific
method. In epidemiology, where variable conditions are the rule, the
repeated observation of associations between covariates by different
investigative teams, in different populations, using different designs
and methods is typically taken as evidence that the association is not
an artifact \cite{1}, for two principal reasons. First, repeated observation adds
quantitative evidence that the association is not due to chance alone;
second, replication across different designs and populations provides
qualitative evidence that the association is not due to uncontrolled
bias affecting a single study. Moreover, accumulated evidence can
provide more accurate estimates of the effect measures of the risk
factor being studied and their uncertainty.

Genetic epidemiology learned the importance of replication the hard way.
Before the advent of genome-wide association (GWA) studies, most
reported geno\-type-phenotype associations failed to replicate. There were
a number of reasons for these conflicting results, including the
following: inappropriate reliance on standard significance thresholds
that did not take the low prior probability of association into account,
small sample sizes, and failure to measure the same variant(s) across
different studies \cite{2,3,4}. In response, the field moved toward more stringent
requirements for reporting associations, explicitly emphasizing
replication \cite{5}. Many high-profile journals now will not publish
genotype-phenotype associations without concrete evidence of
replication \cite{6}.

In this article we review the requirements for replicating associations
discovered via GWA studies in light of recent developments: in
particular, the increasing role of the consortia of multiple GWA
studies. Prospective meta-analysis of multiple genome-wide studies
(conducted by different investigative teams, in different populations,
using different technologies and different designs) can satisfy the
requirement for replication in the context of gene discovery, without
the need to genotype yet more samples in yet further studies, as long as
the combined evidence for association is strong and consistent \cite{7}. This is
an important point, since very large sample sizes are required to
reliably identify common variants with modest effects, and formal
replication of an association---that is, genotyping the initially
discovered genetic variant in a new, completely independent sample of
sufficient size---may be too expensive in terms of time, money and
available samples. Indeed, for some rare diseases (e.g.,
Creutzfeldt--Jakob disease) or relatively uncommon diseases (e.g.,
pancreatic cancer), most if not all samples with\break readily-available DNA
may be genotyped as part of initial GWA studies used at the discovery
stage.

\begin{figure*}

\includegraphics{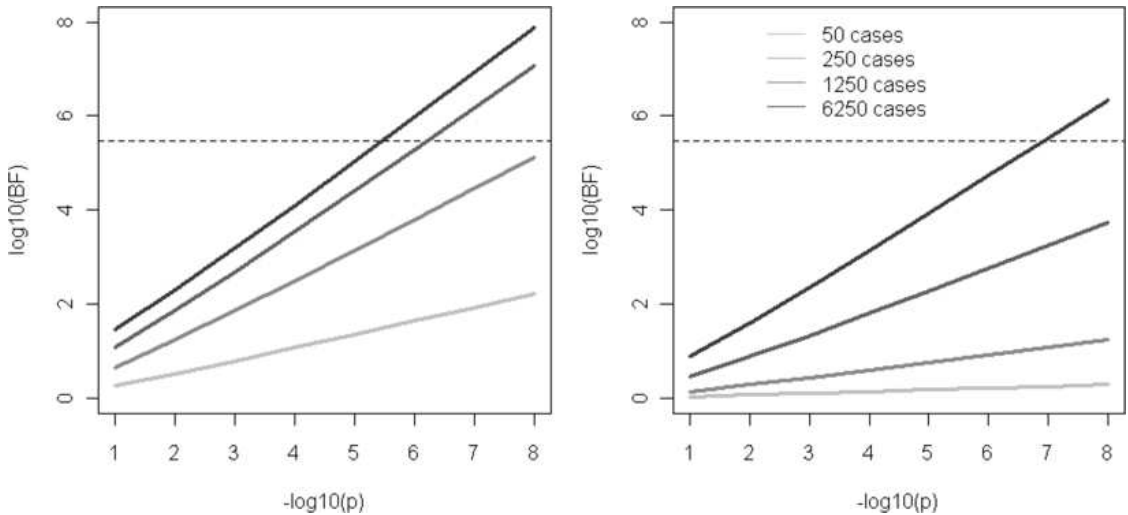}

  \caption{The relationship between the Bayes Factor and $p$-value for
different sample sizes and minor allele frequencies (left panel: minor
allele frequency of 40\%; right panel: 5\%). The dashed line represents
the Bayes Factor necessary to achieve posterior odds in favor of
association of $3:1$ or greater, assuming the prior odds of association
are $1:99\mbox{,}999$. Bayes Factors were calculated for a case-control study
with equal numbers of cases and controls, assuming the expected value of
the absolute value of the log odds ratio is $\operatorname{log}(1.15)$, and assuming a
``spike and smear'' prior. Calculations use equations (4) and (5) from
\cite{13}, with $\sigma^{2} = I_{\beta \beta} ^{-1} -
I_{\alpha \beta} ^{-1} [I_{\alpha \alpha} ^{-1}]^{-1}I_{\beta
\alpha} ^{-1}$, where I is the Fisher information from simple logistic
regression $\operatorname{log}(\mathrm{odds}) = \alpha + \beta G_{\mathrm{additive}}$ calculated
under the null ($\beta =0$).}\label{figure1}
\end{figure*}

We describe the goals of replication and statistical rules of thumb for
distinguishing chance from true associations in the first section of
this article. We then discuss the importance of exact
replication---seeing a consistent association with the same risk allele
using the same analytic methods across multiple studies---and describe
analytic methods for combining evidence across multiple studies, along
with their relative advantages and disadvantages. We close by discussing
why an association may fail to replicate and place replication efforts
in the wider picture of contemporary genetic epidemiology, with its
focus on large-scale collaborations and data sharing.

\section{Goals of Replication} \label{sec:2}

There are two primary reasons replication is essential to confirm
associations discovered via GWA studies: to provide convincing
statistical evidence for association, and to rule out associations due
to biases. Another possible aim of replication is to improve effect
estimation.

\subsection{Convincing Statistical Evidence for Association}

To date most individual GWA studies do not have enough power to detect
true associations at the conservative significance levels necessary to
distinguish false positives from false negatives. This point has
typically been made by referencing the large number of tests conducted
in a GWA study and the consequent severe adjustment of the $p$-value
threshold in order to control experiment-wide Type I error rate.
Empirical estimates of the threshold needed to preserve the genome-wide
Type I error rate in studies of European-ancestry subjects using current
genotyping arrays range from $5\times 10^{-7}$ to $1\times 10^{-8}$ \cite{8,9,10,11}.
These thresholds are different in other populations; for example, they
are even lower in African or African--American samples, due to the
greater genetic diversity in these populations. Even these stringent
thresholds take into account only the complexity of the genetic
architecture, and they do not adjust for the potential complexity of the
phenotypic architecture, that is, when targeting multiple phenotypes.

In the framework of the Bayes theorem, the probability that an observed
association truly exists in the sampled population depends not only on
the observed $p$-value for association, but also on the power to detect
the association (a function of minor allele frequency, effect size and
sample size), the prior probability that the tested variant is
associated with the trait under study, and the anticipated effect size \cite{3,4,12}.
We illustrate this in Figure~\ref{figure1}, where we plot the Bayes Factor for
association (versus no association) as a function of $p$-value, sample
size and minor allele frequency \cite{13}. The Bayes Factor is the ratio of the
probability of the data under the alternative hypothesis (association
with the tested variant) to the probability of the data under the null
hypothesis (no association). (Others define the Bayes Factor as the
inverse of this ratio \cite{13}.) The posterior odds of true association given the
data are equal to the Bayes Factor times the prior odds of association.
In Figure~\ref{figure1} the dashed line represents the Bayes Factor needed to
achieve posterior odds for an association of $3:1$, assuming prior odds of
association of $1:99\mbox{,}999$ (i.e., roughly 100 out of 10,000,000 variants
are truly associated with the studied trait).

Note that all $p$-values are not created equal: for a given $p$-value, the
evidence for association increases with increasing sample size and
depends on risk allele frequency. Increasing overall sample size not
only increases power to detect common risk variants with modest effects,
but it can also increase the credibility of the observed associations.
Differences in credibility by sample size for similar $p$-values are a
consequence of the fact that these calculations take assumptions about
the expected magnitude of the true allelic odds ratio into account. In
particular, they assume that the true allelic odds ratio is unlikely
(probability $<$ 2.1\%) to be smaller than 0.5 or bigger than 2. Since
small $p$-values can only be achieved in small sample sizes if the
estimated effect is large, these results are perceived to be less
credible in this framework.

In other words, the Bayes Factor and thus the credibility of an
association depends explicitly on what we assume for the typical
magnitude of likely genetic effects. For example, if we assume that the
average effect is not an odds ratio of 1.15 as in Figure~\ref{figure1}, but an odds
ratio of $\mathit{OR}_{\mathrm{av}}=1.5$, then the prior odds of association will be
less, because fewer variants---with larger effects than in the
$\mathit{OR}_{\mathrm{av}} =1.15$ scenario---would suffice to explain the genetic
variability. A larger Bayes Factor would be needed to reach a $3:1$
posterior. Moreover, large effects emerging from small studies will be
more credible than in the $\mathit{OR}_{\mathrm{av}} =1.15$ scenario, while very small
effects emerging with similar $p$-values from large studies will be less
credible \cite{13}.

Conversely, if we assume that the average effect is an odds ratio of
$\mathit{OR}_{\mathrm{av}} =1.02$ (consistent with the theory of infinitesimal effects,
each having an almost imperceptible contribution \cite{14}), then the prior odds
of association will be much higher, because a much larger set of
(infinitesimally) associated variants are anticipated and a smaller
Bayes Factor would be needed to reach a $3:1$ posterior. Moreover, large
effects emerging from small studies would be incredible, regardless of
their $p$-value, while very small effects emerging with modest $p$-values
from large studies would provide credible evidence for association.

We should acknowledge that the distribution of effect sizes of true
associations is unknown, and there is no guarantee that they would be
similar for different traits. The difficulty of arriving at the true
causal variants (which may have larger effect sizes than their markers)
adds another layer of complexity. Moreover, given simple power
considerations, it is expected that a large proportion of the large
effects have been identified, while only a small proportion of the
smaller effects and a negligible proportion of the tiny and
infinitesimal effects are already discovered. With these caveats, most
evidence from GWA studies to date is more compatible with the scenarios
of $\mathit{OR}_{\mathrm{av}}$ being in the range of 1.15 \cite{15}, but the 1.02 scenario is not
implausible, and for some traits the 1.5 scenario may be operating, but
we still have not identified the true variants.

Many research groups cannot afford to genotype the large sample sizes
needed to reliably detect genetic markers that are weakly associated
with a trait using a genome-wide platform. This has sparked interest in
multistage designs, where a subset of available samples are genotyped
using the genome-wide platform, and then a subset of the ``most
promising'' markers (typically those with lowest $p$-values) are genotyped
using a custom platform. These designs are reviewed in more detail
elsewhere in this issue \cite{16}. We should note that the primary motivation of
multistage designs is not to increase power by testing fewer hypotheses
in the second stage samples, and hence paying a smaller penalty for
multiple testing at the second stage. Rather the primary goal of
multistage designs is to save genotyping costs or maximize power given a
fixed genotyping budget. If genotyping costs were not an issue, then the
multistage approach is less powerful than simply testing all markers in
the entire available sample \cite{16,17,18}. As genotyping costs decrease, and as more
samples have been genotyped as part of previous GWA analyses,
single-stage analyses become more common \cite{17}.

The appropriate threshold for claiming association depends also on the
context and the relative costs for false positive and false negative
results. For example, re-sequencing a region and conducting in
vivo and in vitro functional studies is quite expensive, and
will require convincing evidence that the observed association is true.
On the other hand, including a region in a predictive genetic risk score
is relatively inexpensive, so a less stringent threshold might suffice.
This approach to replication is intuitively Bayesian (although it need
not use formal Bayesian methods): each successive study serves to update
the prior for association in subsequent studies.

\subsection{Ruling Out Association Due to Artifact}

Even when the initial association is unlikely to be a stochastic
artifact due to multiple testing, it may still be an artifact due to
bias. For common variants, the anticipated effects are modest---for
binary traits, odds ratios smaller than 1.5; for continuous traits,
percent variance explained less than 0.5\%---and very similar in
magnitude to the subtle biases that may affect genetic association
studies---most notably population stratification bias. For this reason,
it is important to see the association in other studies conducted using
a similar (but not identical) study base. In principle, careful design
and analysis should eliminate or greatly reduce bias due to population
stratification in association studies using unrelated individuals \cite{19,20,21}---and,
in practice, these methods have effectively removed some worrying
systematic inflation in association statistics \cite{22}. Family-based designs can
provide additional evidence that an observed association is not due to
population stratification bias, but these designs are not
cost-efficient, and have their own unique sources of bias. For example,
nondifferential genotyping error can inflate Type I error rates in some
family-based analyses, although it does not change the Type I error
rate \cite{23}.

\subsection{Improving Effect Estimates}

Another reason to conduct replication studies is to extend the
generalizablity of the association. It is important to know if the
association exists and has similar magnitude in different environmental
or genetic backgrounds. It is particularly interesting to know how these
associations play out in populations of non-European ancestry,
considering most GWA studies to date have been conducted in
European-ancestry samples. Differences in allele frequencies and local
linkage disequilibrium (LD) patterns across populations present both
challenges and opportunities for replication and fine mapping. On the
one hand, a marker allele that is strongly associated with a trait in
one population may not have a detectable association in another, as the
allele frequency may be smaller or the LD with the (unknown) causal
variant may be much weaker. Thus, initial replication studies should
focus on populations with genetic ancestry similar to that sampled in
the study that first observed the marker-trait association, using the
exact strategy outlined in the next section. Once credible evidence for
this association has been established, replication efforts in other
populations should type not only the marker known to be associated in
the original population, but other markers that ``tag'' common variation
in a region surrounding the marker. For fine mapping, differences in LD
patterns across populations---notably the lower levels of LD in
African-ancestry populations---might lead to refined estimates of the
position of causal variants \cite{24,25}.

Replication may also be useful in identifying a more reliable estimate
of the effect size for the association. Signals selected based on
statistical significance thresholds in underpowered settings are likely
to have (on average) inflated effects due to the winner's curse
phenomenon \cite{26,27,28,30}. Replication should take this into account during the sample
size calculations for the replication efforts; the effect estimate from
the initial study may be inflated, leading to an under-estimate of the
number of subjects needed to reliably detect it \cite{26,27,28,29}. Analytic methods are
available to adjust for winner's-curse bias, but studying the marker in
additional samples (beyond those used to initially identify the marker)
will help produce more unbiased estimates of the genetic effect.
Accurate estimates of marker risks are important (even if the marker is
only a surrogate for the as yet unknown causal variant), as they may be
used for personalized predictive purposes \cite{31,32}.

Finally, when there are several putative association signals in a region
of high LD, dense genotyping in replication studies may help elucidate
whether they represent independent loci, each with its own effect in the
trait, or whether one or all are ``passenger'' markers, which have no
effect conditional on the true underlying causal variant. Detailed
discussion of fine mapping issues is beyond the scope of this review,
but in light of the effort involved, such ``fine mapping'' efforts
should arguably be reserved for loci with credible evidence for
association, for example, loci with markers that have been replicated
exactly, as discussed in the next section~\cite{33}.

\section{Prerequisites for Exact Replication of a Putative
Association from a GWA Study} \label{sec:3}

One of the early difficulties in replicating genetic associations
observed in candidate gene studies was the fact that different groups
would study different markers in the same region. Because the LD among
these markers was poorly understood, results from multiple studies could
increase rather than decrease confusion. The initial study may have seen
an association with SNP~A, but the second study did not genotype that
SNP, and instead saw an association with SNP B, which was not genotyped
in the original study. As the number of SNPs typed per region increased,
``moving the goalposts'' in this fashion contributed to the problem of
persistent false positives in the candidate gene literature; by chance,
some SNP in the region (not necessarily the SNP that was statistically
significant in other studies) would have $p<0.05$, and this would be
(incorrectly) proclaimed replication \cite{34}. In response to this problem,
guidelines for replication in genetic association studies now call for
exact replication. The same marker---or, if technical difficulties
preclude this, a perfect or near-perfect proxy for the original
marker---should be genotyped across all studies and analyzed using the
same genetic model. In this section we discuss prerequisites for exact
replication. We use the term ``exact replication'' cautiously,
recognizing that this is an unattainable goal in epidemiology (e.g.,
studies conducted by different investigators at different times, let
alone places, will sample from different populations) and that in some
sense it is the ``inexactness'' of replication studies that increases
credibility of the observed association (it is less likely to be an
artifact due to a bias that is unique to the initial study). We use the
term to emphasize the danger of ``moving the goalposts'' so far that
claims of replication carry little weight.

\subsection{Test the Same Marker}

This should be done preferably by directly genotyping this marker.
Currently-available imputation methods are powerful and quite accurate
for filling in information on missing common SNPs \cite{35,36,37,38,39}. Even then, further
confirmation by direct genotyping would be very useful. (In fact, to
rule out technical artifact, some have argued that an associated SNP
should be genotyped using two different genotyping technologies, or that
a second SNP in the region that is in [near-] perfect LD with the
associated SNP be genotyped \cite{5}.) Great caution is needed when
``replicating'' an association by finding an association with a
(different) nearby marker: if the new marker does not have perfect or
almost perfect LD with the previously discovered one, this cannot be
considered replication. Moreover, even for markers with seemingly
perfect LD in a given sample, the LD may be far less than perfect in a
different population and it may break completely in populations of
different ancestry. When a panel of markers spanning the whole locus is
pursued (e.g., after resequencing and fine mapping), different markers
and haplotypes may be found to be associated in different populations.
Evidence from different markers and haplotypes should not be combined in
the same meta-analysis. The consistency of each association can be
formally assessed separately (see the section on statistical
heterogeneity).

\subsection{Use the Same Analytic Methods}

If the initial results found an increased risk per copy of, say, the A
allele (additive model), then a significant increased risk for carriers
of the T allele (dominant model, in other direction) does not constitute
replication. It is in principle possible that the direction of
association can change due to differences in linkage disequilibrium
across study populations. However, this ``flip flop'' phenomenon can
occur only in very specific situations that are unlikely when the study
populations have similar continental ancestry \cite{40}. The burden of proof is on
investigators to show evidence for how difference in LD in their study
populations could produce a ``flip flop'' if they wish to claim
replication, even though different alleles are associated with risk.
Merely citing the possibility of ``flip-flopping'' does not suffice.

Other analytical options include the statistical\break model (e.g., for a
binary outcome, whether it is\break treated as simply yes/no or the
time-to-event is also taken into account), the use of any covariates
(e.g. for age, gender or topic-specific variables) and the use of
corrections for relatedness. Usually, the impact of these options is not
major, but it can make a difference for borderline associations which
may seem to pass or not pass a desired $p$-value threshold. This means
that both for GWA studies and subsequent investigations, one should
carefully report the methods in sufficient detail so they can be
independently replicated by other researchers \cite{41}.

Modeling can have a much more profound impact in more complex
associations than go beyond single markers, for example, with approaches
that try to model dozens and hundreds of gene variants that form a
``pathway'' \cite{42,43}. Such complex models may be built by MDR, kernel machines,
stepwise logistic regression or a diversity of other methods and it is
important for the replication process to use the same exact steps as the
model building. Even then, because these models are so flexible, it is
unclear whether a ``significant'' finding in a second data set
constitutes replication; the association may be driven by different sets
of SNPs in the different studies. Researchers who conduct complex
model-selection/model-building analyses should report their ``final''
model in as much detail as possible, so other investigators can judge
the fit of that model in other data sets.

\subsection{Try to Use the Same Phenotype}

For many traits, phenotype definitions may vary considerably across
studies, or there may be many different options for defining the
phenotypes of interest within each study. Some of this variability is
unavoidable and results from differences in measurement protocols across
studies. For example, disease may be self-reported in some studies or
clinician-diagnosed in others; waist:hip ratios may be self-reported or
measured in a clinic, using different operational definitions of
``waist''; etc. Characteristics of studied phenotype may also differ
across studies: for example, because of the widespread use of
Prostate-Specific Antigen (PSA) screening in the United States since the
early 1990s, the proportion of early-stage prostate cancer cases in the
US is higher than in Europe, where PSA screening is not as common. In
the context of a prospective meta-analysis, study investigators can
discuss these issues and reach consensus on how to define phenotype so
as to maximize relevant information while ensuring as many studies can
provide data as possible. In general, there is a trade off between more
accurate (but more expensive and perhaps more invasive) measurements on
fewer people and less accurate (but cheaper) measurements on more
people. For example, although the Fagerstom Test may be a ``gold
standard'' measure of nicotine dependence, currently only a few studies
with available genome-wide genotype data have collected data on this
test; on the other hand, many studies have collected information about
the number of cigarettes smoked per day (a component of the Fagerstrom
score) \cite{44}. To maximize sample size, investigators may agree to analyze
cigarettes per day (which then raises further issues such as what scale
to use, whether and how to transform the raw data, how to reconcile
continuous with categorical data, etc.). Prospective meta-analyses for
height, BMI and fasting glucose have dealt with the issue of phenotype
harmonization in a trait-by-trait basis \cite{45,46,47,48}. Other consortia and projects
such as the Public Population Project in Genomics
(\url{http://www.p3gconsortium.org/})\break and PhenX (\href{http://www.phenx.org}{www.phenx.org}) aim to
facilitate broad collaboration among existing and future genome-wide
association studies by making recommendations for standard phenotyping
protocols for many diseases and traits. Still, despite best efforts to
harmonize measures, some measurement differences\break across studies will
persist, and investigators should be aware of these as possible sources
of heterogeneity (see Sections~\ref{sec:4} and~\ref{sec:5}).

Requiring that replication studies use the same phenotype definition
used in the initial study also helps avoid false positives due to ``data
dredging,'' the temptation to generate small $p$-values by testing many
different traits (different case subtypes, continuous traits dichotomize
using different, arbitrary cut points, etc.) \cite{4}. When many phenotypes or
phenotype definitions and analyses are used, there should be a penalty
for multiple testing. Applying this penalty is not always
straightforward, given that most of the phenotypes and analyses are
usually correlated or even highly correlated. However, the danger exists
for an association to be claimed replicated, after searching through
repeated modifications of the phenotypes and analyses thereof. A $p$-value
that has been obtained through such an iterative searching path is not
the same as one that was obtained from a single main analysis of a
single phenotype.

\section{Replication Methods and Presentation of Results} \label{sec:4}

\subsection{Statistical Heterogeneity Across Datasets}

There are several tests and metrics of between-dataset heterogeneity,
borrowed from applications of meta-analysis in other fields. The most
popular are Cochran's $Q$ test of homogeneity \cite{49}, the $I^{2}$ metric
[obtained by ($Q\mbox{-degrees of freedom})/Q$] and the be\-tween-study variance
estimator $\tau^{2}$ \cite{50}. There are shortcomings to all of them \cite{51}. The $Q$ test
is underpowered in the common situation where there are few datasets and
may be overpowered when there are many, large datasets. There are now
readily-available approaches that can be used to compute the power of
the $Q$ test to detect a given tau-squared \cite{52}. When the~$Q$ test is
underpowered, the $I^{2}$ metric has large uncertainty and this can be
readily visualized by computing its 95\% confidence intervals \cite{53}.
Similarly, estimates of $\tau^{2}$ may have large uncertainty. One
potentially useful approach may be to estimate the magnitude of
between-study variability compared with the observed effect size
$\theta$, that is, $h=\tau /\theta$. For a small effect size, even small
$\tau^{2}$ may question the generalizability of the conclusion that
there is an association across all datasets. This conclusion would not
be as easily challenged in the presence of a large effect size.

Some other caveats should be mentioned. The winner's curse in the
magnitude of the effect in the discovery phase may introduce spuriously
inflated heterogeneity, when the discovery data are combined with
subsequent replication studies. In such two-stage approaches,
between-study heterogeneity\break should best be estimated excluding the
discovery data. Conversely, if all datasets are measured with
genome-wide platforms and GWA scan meta-analysis is performed in all
gene variants, this is no longer an issue. In fact, if the GWA scan
meta-analysis uses random effects (see below), the emerging top hits
from the GWA scan meta-analysis are likely to have, on average, deflated
observed heterogeneity compared with the true heterogeneity. This is
because underestimation of the between-study heterogeneity favors a
variant to come to the top of the list, since it does not get penalized
by wider confidence intervals in the random effects setting.

\begin{figure}

\includegraphics{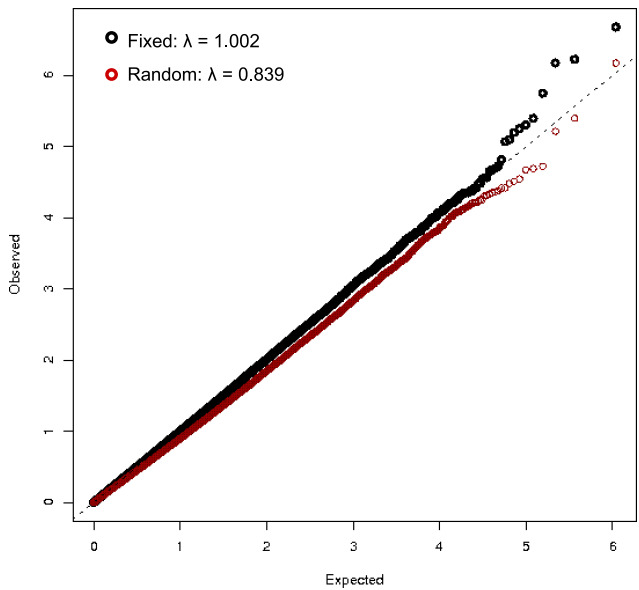}

\caption{Quantile-quantile plots for fixed-effect and random-effect
meta-analyses of the 13 studies in the initial PanScan genome-wide
association study of pancreatic cancer. The genomic control inflation
factors $\lambda_{\mathrm{GC}}$ for the fixed-effect and random effect
analyses were 0.84 and 1.00, respectively. $\lambda_{\mathrm{GC}}$ was
calculated as the median observed chi-squared test statistic divided by
the median of a chi-squared distribution with one degree of freedom.}\label{figure2}
\end{figure}

However, we caution that when the number of studies is relatively small,
association tests based on random-effects meta-analysis may be deflated,
as the between-study variance $\tau^{2}$ will be poorly estimated. This
is illustrated in Figure~\ref{figure2}, which shows quantile-quantile plots for
fixed-effect and random-effects meta-analyses of data from PanScan
collaboration, which involves 13 studies in the initial GWAS scan. For
the random effects analysis, the genomic-control ``inflation factor'' is
in this case more aptly named a ``deflation factor'':
$\lambda_{\mathrm{GC}}=0.84$, indicating that the random effects $p$-values are
larger than expected under the assumption that the vast majority of SNPs
are not associated with pancreatic cancer. Fixed-effect meta analysis is
arguably more appropriate as an initial screening test for associated
markers, although because fixed-effect analysis can be highly
significant when only one (relatively large) study shows evidence for
association, analyses that incorporate effect heterogeneity such as
random effects meta-analysis should be reported for highly significant
markers from fixed-effect analyses.

Finally, lack of demonstrable heterogeneity may be perceived as a
criterion of credible replication \cite{54}. However, one should note that tests
and measures of heterogeneity address whether effect sizes across
different datasets vary, not whether they are consistently on the same
side of the null. Dataset-specific effects could vary a lot, but they
may all still point to the same direction of effect. Given the potential
diversity of LD structure across populations, and differences in
phenotype definitions and measurements across studies, between-study
heterogeneity should not dismiss an association because the effect sizes
are not consistent, if the evidence for rejection of the null hypothesis
is strong.

\subsection{Models for Synthesis of Data from Multiple Replication
Studies}

Data across studies can be combined at the level of either $p$-values
(probability pooler methods) or effect sizes (effect size
meta-analysis) \cite{36,55,56}. When $p$-values are combined, at a minimum one should take
into account also the direction of effects, but the magnitude of the
effects is not taken into account. When effect sizes are used, there are
several models that can be used, depending on whether between-study
heterogeneity is taken into account or not, and if the former, how this
is done. In general, fixed effects approaches that ignore between-study
heterogeneity are better powered than random effects approaches and thus
more efficient for discovery purposes. However, there is a trade-off for
increased chances of false-positives. For effect estimation and
predicting what effects might be expected in future similar populations,
random effects are intuitively superior in capturing better the extent
of the uncertainty. Commonly, random effects are estimated with a 95\%
CI that captures the uncertainty about the mean effect, but ideally one
should also examine the uncertainty of the distribution of effects
across populations. This is provided by the prediction interval. An
approximate $(1-\alpha)\%$ prediction interval for the effect in an
unspecified study can be obtained from the estimate of the mean
effect $\hat{\mu}$ , its estimated standard error and the estimate of the
between-study variance $\hat{\tau}^{2}$ by
\[
\hat{\mu} \pm t_{k - 2}^{\alpha} \sqrt{} \{ \hat{\tau} ^{2} +
\widehat{\operatorname{SE}}(\hat{\mu} )^{2}\},
\]
where $t_{k - 2}^{\alpha}$ is the $100(1-\alpha /2)\%$ percentile of the
$t$-distribution with $k-2$ degrees of freedom \cite{57}. It becomes implicit that
when an association has been probed in only a few datasets, then the
prediction interval will be wider than the respective confidence
interval, even if there is no demonstrable between-study variance (i.e.,
$\tau^{2}=0)$. Table~\ref{tab1} summarizes some issues that arise in selecting,
interpreting and comparing the properties and results of various
commonly used meta-analyses methods.

\section{Reasons for Nonreplication } \label{sec:5}

\begin{quote}

\textit{[T]here are often two or more hypotheses which account for all
the known facts on some subject, and although, in such cases, men [sic]
of science endeavour to find facts which will rule out all the
hypotheses except one, there is no reason why they should always
succeed.} --- Bertram Russell \cite{58}

\end{quote}

A variant observed to be associated with a trait in an initial GWA may
not be associated with the trait in subsequent studies, even though the
original association was (nearly) ``genome-wide significant.'' There are
a number of potential reasons for this nonreplication.

\begin{table*}
\caption{Different methods for meta-analysis in the genome-wide
association setting}\label{tab1}
\begin{tabular*}{380pt}{@{\extracolsep{4in minus 4in}}p{118pt}@{\hspace*{22pt} }lp{62pt}p{62pt}@{}}
\hline
\textbf{Issues and caveats} & \textbf{\textit{P}-value}  & \multicolumn{2}{l@{}}{\textbf{Effect size meta-analysis}}\\
 & \textbf{meta-analysis} &  & \\
\ccline{3-4}
 &  & \textbf{Fixed effects} & \textbf{Random effects}\\
 \hline
Direction of effect is considered & In some methods & Yes & Yes\\
Effect size is considered & No & Yes & Yes\\
Summary $p$-value is obtained & Yes & Yes & Yes\\
Summary effect is obtained & No & Yes & Yes\\
\mbox{Summary result can be} \mbox{converted to credibility based}
\mbox{on priors for the anticipated} \mbox{effect sizes} & No & Yes & Yes\\
\mbox{Between-study heterogeneity} \mbox{can be taken into account} & No & No & Yes\\
\mbox{Between-study heterogeneity} \mbox{can be estimated/tested} & No & Yes & Yes\\
\mbox{Consensus on if/how datasets} should be weighted & No & Yes & Yes\\
Commonly used weights & None, SQRT(N), N & Inverse variance & Inverse variance\\
\mbox{Prior assumptions on the effect} \mbox{size can be used} & No & \mbox{In Bayesian} meta-analysis & \mbox{In Bayesian} meta-analysis\\
\mbox{Prior uncertainty on} \mbox{heterogeneity can be} \mbox{accommodated} & No & No & \mbox{In Bayesian} meta-analysis\\
Prior uncertainty on the genetic model can be accommodated & No & \mbox{In Bayesian} \mbox{M-A} & \mbox{In Bayesian} meta-analysis\\
\mbox{Normality assumptions typically} made within each study & Yes & Yes & Yes\\
\mbox{Normality assumptions within} each study easily testable & Yes, rarely done & Yes, rarely done & \\
\mbox{Normality assumptions for} \mbox{distribution of effects across} studies easily testable & No effects assumed & \mbox{Single common} \mbox{effect assumed} (assumption \mbox{may be visibly} wrong) & \mbox{Not easily} testable\\
\mbox{Heavy-tail alternative methods} exist & No & Yes, rarely used & Yes, rarely used\\
\mbox{Use with uncommon alleles} \mbox{(small genotype groups, or}
even zero allele counts in $2\times2$ tables) & Need to use exact methods & Quite robust & \mbox{Between-study} variance \mbox{estimation} unstable\\
Power for discovery & Good & Good & Less than others\\
\mbox{False-positives from single} \mbox{biased dataset} & Susceptible & Susceptible & Less susceptible\\
\mbox{False-positives when evidence} \mbox{from small studies is most} biased & Susceptible & Susceptible & More susceptible\\
\mbox{False-positives when evidence} \mbox{from large studies is most biased} & Susceptible & Susceptible & Less susceptible\\
\mbox{Can predict range of effect sizes} \mbox{in future similar populations} & No & Too narrow confidence intervals & Appropriate \mbox{with predictive} \mbox{intervals}\\
\mbox{Can convey uncertainty for} \mbox{practical applications (e.g., to} be used in clinical prediction test) & Useless & Inappropriate & Most appropriate \mbox{with prediction} intervals\\
\hline
\end{tabular*}
\end{table*}

\begin{longlist}
\item[(a)] The original observation was a false positive due to sampling
error. This is the default explanation, \mbox{until} proven otherwise. This is
more likely for associations that were not (or just barely)
``genome-wide significant'' than for observations that were extremely
statistically significant.

\item[(b)] The follow-up study had insufficient power. This problem can
be avoided by ensuring the follow-up study is large enough to reliably
detect the observed effect (after accounting for inflation due to
``winner's curse'') \cite{26,27,28,29}. Moreover, if we consider the cumulative evidence
(both the original data plus the follow-up data) as an updated
meta-analysis, the cumulative evidence may still pass\break genome-wide
significance or a sufficient Bayes Factor threshold, even though the
follow-up data are not formally (highly) significant, when seen in
isolation.

\item[(c)] The genotypic coding used in the initial study may not
accurately reflect the true underlying association, leading to a loss of
power. Ideally the follow-up study should be well powered to detect
associations based on different genetic models (e.g., recessive,
dominant) that are consistent with the results observed in the first
study.

\item[(d)] The variant may be a poor marker for the trait due to
differences in linkage-disequilibrium\break structure between the studies.
This is more likely if the study populations have different ethnic
backgrounds. When discussing this as a possible reason for
nonreplication, investigators should make a good-faith effort to provide
empirical data on how\break linkage-disequilibrium patterns differ (e.g.,
using\break HapMap data) and how these differences would lead to
inconsistencies across studies.

\item[(e)] Differences in design or trait definition may lead to
inconsistencies. See Sections~\ref{sec6.1} and~\ref{sec6.2} for examples of how different
matching or ascertainment schemes can affect estimates of marker-trait
association. Again, when citing this as a reason for nonreplication,
investigators should as far as possible present arguments for the
likelihood and magnitude of differences due to design or measurement
differences.

\item[(f)] The absence of an association in the subsequent studies may be
due to true etiologic heterogeneity. Sometimes, this may be driven by
gene--gene or gene--environment interaction. If cases in the original
study were required to have a family history of disease, for example, or
required to have a relatively rare exposure profile (e.g., male lifetime
never smokers), then subsequent studies that do not impose these
restrictions may not see the association, if the association is
restricted to subgroups with a particular genetic or exposure
background. However, to date, gene--gene and gene--environment
interactions have been notoriously difficult to document robustly.
\end{longlist}

For the last three explanations, it is useful to clarify if the
explanation was offered a posteriori after observing the inconsistent
results in different studies. Post hoc explanations for subgroup
differences, interactions and effect modification may be overfit to the
observed data and may require further prospective replication in further
datasets before they can be relied upon.

\section{The Wider Picture of Replication Efforts: Consortia, Data
Availability and~Field~Synopses} \label{sec:6}

With the recent successes of GWA studies, the field has realized that
increasingly large sample sizes are required to identify and replicate
the increasingly small effect sizes at common variants that remain
undetected. Even wider networks will be required to facilitate the study
of variation at the lower end of the frequency spectrum (be it single
base changes, copy number variants or otherwise). Collaboration and data
sharing are invaluable tools in achieving the necessary sample sizes
for\break
well-powered replication studies. The past few years have witnessed a
rapid rise in international consortium formation and collaboration has
taken a most prominent role in conducting research. Consortia allow
investigators to make some design choices up front (if only deciding
which SNPs to attempt to replicate), and to work together to harmonize
phenotypes and analyses \cite{7}. Several examples of notable successes of
consortium-coordinated efforts have started to emerge in the literature \cite{47,59,60,61,62}.

\begin{table*}
\caption{Cumulative power to detect association ($\alpha =5\times10^{-8}$)
at a risk allele with frequency 0.20 and 0.40, and allelic odds ratios
of 1.1 and 1.2, given sample sizes for the WTCCC, DGI and FUSION
studies}
\label{tab3}
\begin{tabular*}{380pt}{@{\extracolsep{\fill}}lccccc@{}}
\hline
\textbf{Studies} & \textbf{Risk allele}  & \textbf{Allelic}
& \textbf{Cumulative \textit{n}}   & \textbf{Cumulative \textit{n}}  & \textbf{Power}\\
&\textbf{frequency}&\textbf{odds} \textbf{ratio}&\textbf{cases}&\textbf{controls}\\
\hline
WTCCC & 0.20 & 1.10 & 1924 & 2938 & 0.0002\\
WTCCC${}+{}$DGI & 0.20 & 1.10 & 3388 & 4405 & 0.0011\\
WTCCC${}+{}$DGI${}+{}$FUSION & 0.20 & 1.10 & 4549 & 5579 & 0.0033\\
WTCCC & 0.40 & 1.10 & 1924 & 2938 & 0.0007\\
WTCCC${}+{}$DGI & 0.40 & 1.10 & 3388 & 4405 & 0.0054\\
WTCCC${}+{}$DGI${}+{}$FUSION & 0.40 & 1.10 & 4549 & 5579 & 0.0166\\
WTCCC & 0.20 & 1.20 & 1924 & 2938 & 0.0333\\
WTCCC${}+{}$DGI & 0.20 & 1.20 & 3388 & 4405 & 0.2078\\
WTCCC${}+{}$DGI${}+{}$FUSION & 0.20 & 1.20 & 4549 & 5579 & 0.4426\\
WTCCC & 0.40 & 1.20 & 1924 & 2938 & 0.1336\\
WTCCC${}+{}$DGI & 0.40 & 1.20 & 3388 & 4405 & 0.5468\\
WTCCC${}+{}$DGI${}+{}$FUSION & 0.40 & 1.20 & 4549 & 5579 & 0.8219\\
\hline
\end{tabular*}
\end{table*}

In silico replication of association signals has been further
facilitated by initiatives making genetic association study results
and/or raw data publicly available (or available through application to
an access committee), for example, the Wellcome Trust Case
Control Consortium (\href{http://www.wtccc.org.uk}{www.wtccc.org.uk}),
dbGAP\break
(\href{http://www.ncbi.nlm.nih.gov/sites/entrez?db=gap}{http://www.ncbi.nlm.nih.gov/sites/entrez?db=gap})
and the European
Genotype Archive (EGA,\break \url{http://www.ebi.ac.uk/ega}). Several emerging
considerations, for example, with respect to the\break anonymity of data \cite{63},
avenues for communication between primary investigators and secondary
users to facilitate a better understanding of the datasets and their
appropriate uses, and suitable accreditation of involved parties,
require resolution in order to optimize the use of publicly available
raw data.

Replication undoubtedly constitutes an evolving practice. The need to
incorporate new data arising from further GWA scans, other replication
studies, meta-analyses or all of the above leads to the emerging
paradigm of conglomerate analyses. Field synopses, for example, are
efforts to integrate data from diverse sources (GWA studies, consortia,
single published studies) in the published literature and to make them
publicly available in electronic databases that can be updatable.
Examples include the field synopses on Alzheimer's disease (AlzGene
database), schizophrenia (SzGene database) and\break DNA repair genes \cite{64,65}. The
results of the meta-analyses on the accumulated data can then also be
graded for their epidemiological credibility, for example, as proposed
by the Venice criteria \cite{54}.

\subsection{Example from the Field of Type 2 Diabetes}\label{sec6.1}

Researchers in the field of Type 2 diabetes (T2D) genetics were among
the first to lead the way in distributed collaborative networks,
exemplified by early efforts such as the International Type~2 Diabetes
Linkage Analysis Consortium and the International Type~2 Diabetes 1q
Consortium \cite{66,67,68}. The advent of GWA scans was met by pre-publication data
sharing between three large-scale studies, the WTCCC, DGI and FUSION
scans \cite{9,69,70,71}, leading to the formation of the DIAGRAM Consortium (Diabetes
Genetics Replication and Meta-analysis). By exchanging information on
top signals, the three studies obtained in silico replication
of individual scan findings and then further pursued de novo
replication in additional sets of independent samples. This endeavor
additionally highlighted examples of statistical heterogeneity across
the studies, notably with respect to one of the\break WTCCC study's strongest
signals, residing within the \textit{FTO} gene \cite{72}. This inconsistency in
observed associations could be ascribed to study design and,
specifically, to matching cases and controls for BMI (DGI study). The
\textit{FTO} signal was quickly identified as the first robustly
replicating association with obesity, mediating its effect on T2D
through BMI. A truly genome-wide meta-analysis of the three scans
ensued, with large-scale replication efforts in independent datasets of
T2D cases and controls, all of European origin. This effort led to the
identification of further novel T2D susceptibility loci \cite{59}.
\mbox{Table~\ref{tab3}}
demonstrates the gains in power afforded by increasing sample size from
a single scan to the synthesis of all three studies for a realistic
common complex disease susceptibility locus.

\subsection{Anthropometrics and the Analysis of ``Secondary
Traits''}\label{sec6.2}

The meta-analyses of body mass index and height conducted by the Genetic
Investigation of ANthropometric Traits (GIANT) consortium raised
additional issues \cite{45,47,73}. Specifically, unlike the diabetes consortia, where
each participating study was designed with diabetes as its primary
outcome, the studies involved in \mbox{GIANT} were not originally designed to
study determinants of BMI and height, rather they were originally
case-control studies of diabetes, prostate and breast cancers, and other
diseases \cite{74,75}. In principle, if the studied trait is associated with disease
risk, then conditioning on case-control status can create a spurious
association between a marker and the trait. In practice, only a small
number of markers will have an inflated Type I error rate---namely,
those markers that are associated with disease risk but not directly
with the secondary trait---and the magnitude of the inflation depends on
both the strength of the association between the secondary trait and
disease (which could be modest or controversial, as in the case of
smoking and breast or prostate cancer, or quite strong, as in the case
of BMI and T2D or smoking and lung cancer) and the strength of the
association between the marker and disease (typically relatively weak) \cite{75,76}.
Moreover, the risk of false positives may be further ameliorated by
diversity of designs among the participating studies---some may have
originally been case-control studies of different diseases, others may
have been cohort or cross-sectional studies. Although there are analytic
methods that can eliminate spurious association or bias due to
case-control ascertainment in particular situations and under particular
assumptions \cite{74,76}, these should not replace careful consideration of potential
biases and evaluation of heterogeneity in effect measures across studies
with different designs.

\section*{Acknowledgments}

EZ is supported by the Wellcome Trust\break (WT088885/Z/09/Z). We thank Mandy
van Hoek for contributing to power calculations. Scientific support for
this project was provided through the Tufts Clinical and Translational
Science Institute (Tufts CTSI) under funding from the National Institute
of Health/National Center for Research Resources (UL1 RR025752). Points
of view or opinions in this paper are those of the authors and do not
necessarily represent the official position or policies of the Tufts
CTSI.

\end{document}